\documentclass{article}[12pt]
\usepackage{amsfonts}
\usepackage{amsmath}
\usepackage{amsthm}
\usepackage{enumerate}
\usepackage[british]{babel}
\usepackage[T1]{fontenc}
\usepackage{boxedminipage}
\usepackage{array}

\newcommand{\cF}{\mathcal F}
\newcommand{\cA}{\mathcal A}
\newcommand{\cW}{\mathcal W}
\newcommand{\GF}{\mathrm{GF}}
\newcommand{\PG}{\mathrm{PG}}
\newcommand{\cC}{\mathcal C}
\newcommand{\cH}{\mathcal H}
\newcommand{\cL}{\mathcal L}
\newcommand{\tr}{\mathrm{T}\,}
\newtheorem{lemma}{Lemma}
\newtheorem{theorem}[lemma]{Theorem}
\theoremstyle{remark}
\newtheorem{remark}{Remark}

\begin{document}
\title{Construction of a $3$--dimensional MDS-code\\
\vspace*{0.5cm}\small{\sc Dedicated to the centenary of the birth of
\\ Ferenc K\'arteszi (1907-1989)\vspace*{1cm}}}
\author{A. Aguglia and L. Giuzzi\thanks{Research supported by  the Italian
    Ministry MIUR, Strutture geometriche, combinatoria e loro
    applicazioni.}}
\date{}

\maketitle

\begin{abstract}
In this paper, we describe a procedure for constructing $q$--ary
$[N,3,N-2]$--MDS codes, of length $N\leq q+1$ (for $q$ odd) or
$N\leq q+2$ (for $q$ even), using a set of non--degenerate Hermitian
forms in $PG(2,q^2)$.
\end{abstract}

\section{Introduction}

The well--known Singleton bound states that the cardinality $M$ of
a code of length $N$ with minimum distance $d$ over a $q$--ary
alphabet always satisfies
\begin{equation}\label{Sing} M\leq q^{N-d+1};
\end{equation}see \cite{S1}. Codes attaining the  bound
are called \emph{maximum distance
separable codes}, or \emph{MDS} codes for short.

Interesting  families of maximum distance separable  codes arise
from geometric and combinatorial objects embedded in a finite
projective spaces. In particular linear $[N,k,N-k+1]$--MDS codes,
with $k\geq 3$,  and $N$--arcs in $PG(k-1,q)$ are equivalent
objects; see \cite{AK}.
%The so called Main Conjecture for MDS codes states that for $q>k$,
%the maximum length $N$ of a code is
%\[ M(k,q)=\begin{cases}
 % q+2 & \mbox{for $q$ even and either $k=3$ or $k=q-1$} \\
 % q+1 & \mbox{in all other cases}
  %\end{cases}
%\]

%Here we present a procedure  depending on Hermitian forms in
%$PG(2,q^2)$, that provides $3$--dimensional MDS codes.
A general method for constructing a $q$--ary code is to take $N$
multivariate polynomials $f_1,\ldots,f_N$ defined over a suitable
subset $\cW\subseteq\GF(q)^{n}$ and consider the set $\cC$ given
by
\[\cC=\{(f_1(x),\ldots,f_N (x))|x\in \cW\}.\]
In this paper, we deal with the case  $|\cW|=q^t$ and also assume
that the \emph{evaluation function}
\[ \Theta:\begin{cases}
  \cW\mapsto\cC \\
  x \mapsto (f_1(x),f_2(x),\ldots,f_N(x))
  \end{cases}
\]
 is injective.

If $\cC$ attains the Singleton bound  then the restrictions of
all the codewords to any
given $t=N-d+1$ places must all be different, namely in
any $t$ positions all possible vectors occur exactly once. This
means that a necessary condition for $\cC$ to be MDS is that any
$t$ of the varieties $V(f_m)$ for $m=1,\ldots,N$ meet in exactly
one point in $\cW$. Here $V(f)$ denotes the algebraic variety
associated to $f$.
\par
Applying the above procedure to a set of non--degenerate Hermitian
forms in $PG(2,q^2)$ we construct some $q$--ary $[N,3,N-2]$--MDS
codes, of length $N\leq q+1$ (for $q$ odd) or $N\leq q+2$ (for $q$
even). The codes thus obtained can also be  represented by sets of
points in $\PG(3,q)$; this representation is used in Section
\ref{decod} in order to devise an algebraic decoding procedure,
based upon polynomial factorisation; see \cite{Su}.

% The same
%decoding procedure could also be adapted to correct errors
%``beyond the bound'', that is to correct, under favourable
%circumstances, more than $e=\lfloor(d-1)/2\rfloor$ errors in a
%received word.

\section{Preliminaries}
Let $\cA$ be a set containing $q$ elements. For any integer $N\geq
1$, the function $d_H:\cA^N\times\cA^N\mapsto{\mathbb N}$
given by
\[ d_H(\mathbf{x},\mathbf{y})=|\{i: x_i\neq y_i\}|, \]
is a metric on $\cA^N$. This function is called  the \emph{Hamming
distance} on $\cA^N$. A $q$--ary $(N,M,d)$--code $\cC$ over the
alphabet $\cA$ is just a collection of $M$ elements of $\cA^N$
such that any two of them are either the same or at Hamming
distance at least $d$; see \cite{LG,E}.
 The elements of $\cC$ are called \emph{codewords} whereas
 the integers $d$ and $N$ are respectively
 the \emph{minimum distance} and the \emph{length}
 of $\cC$.
 \par
If $\cA=GF(q)$ and $\cC$ is a $k$--dimensional vector subspace of
$GF(q)^N$, then $\cC$ is said to be a \emph{linear} $[N,k,d]$--code.
Under several communication models, it is assumed that a received
word $\mathbf{r}$ should be decoded as the codeword $\mathbf{c}\in\cC$
which is nearest to $\mathbf{r}$ according to the Hamming distance;
this is the so--called maximum likelihood decoding.
Under these assumptions
the following theorem, see \cite{LG,E}, provides a basic bound on
the guaranteed error correction capability of a code.
\begin{theorem}
If $\cC$ is a code of minimum distance $d$, then $\cC$ can always
either detect up to $d-1$ errors or
correct $e=\lfloor(d-1)/2\rfloor$ errors.
\end{theorem}
Observe that the theorem does not state
that it is not possible to decode a word  more than $e$
errors happened, but just that in this case it is possible
that the correction fails. Managing to recover from more than $e$ errors
for some given received codewords is called ``correcting beyond the
bound''.

The \emph{weight} of an element ${\mathbf x}\in GF(q)^N$ is
the number of non--zero components $x_i$ of $\mathbf{x}$.
For a linear code the minimum distance $d$  equals  the minimum weight of
the non--zero codewords.
\par
The parameters of a code are not independent; in general it is
difficult to determine the maximum number of words a code of
prescribed length $N$ and minimum distance $d$ may contain.
%several limitations are known, one of the most basic limitations
%on the number of codewords  is the Singleton bound, see \cite{S1},

%Any code $\cC$ attaining this bound is called \emph{maximum distance
%separable} or, in short, \emph{MDS}.

Observe that, for any arbitrary linear $[N,k,d]$--code, condition
\eqref{Sing} may be rewritten as
\begin{equation}\label{Sing1}
d\leq N-k+1;
\end{equation}
thus $\cC$ is a linear MDS code if and only if equality holds in
\eqref{Sing1}.

In Section \ref{s:cons} we shall make extensive use of some
non--degenerate Hermitian forms in $PG(2,q^2)$.

Consider the projective space $PG(d, q^2)$ and let $V$ be the
underlying vector space of dimension $d + 1$.  A \emph{sesquilinear
Hermitian form} is a map
\[h: V \times V \longrightarrow GF(q^2) \]
additive in both components and satisfying
\[h(k\mathbf{v}, l\mathbf{w}) = kl^qh(\mathbf{v},\mathbf{w})\]
for all $\mathbf{v},\mathbf{w} \in V$  and $k, l \in  GF(q^2)$.
The form is \emph{degenerate} if and only if the subspace
$\{\mathbf{v} | \ h(\mathbf{v},\mathbf{w}) = 0 \  \forall
\mathbf{w}\in V \}$, the \emph{radical} of $h$, is different from
$\{\mathbf{0}\}$. Given a sesquilinear Hermitian form $h$, the
associated Hermitian variety $\cH$ is the set of all points of
$PG(d,q^2)$ such that $\{ <\mathbf{v}>
 | \mathbf{0}\neq \mathbf{v} \in V,  \ h(\mathbf{v},\mathbf{v}) = 0\}$.
 The variety $\cH$ is \emph{degenerate} if $h$ is degenerate;
 non--degenerate
 otherwise.
If $h$ is a sesquilinear Hermitian form in $PG(d,q^2)$ then
 the map $F : V \longrightarrow GF(q)$
defined by \[F(\mathbf{v}) = h(\mathbf{v},\mathbf{v}),\] is called
\emph{the Hermitian form on $V$ associated to $h$}. The Hermitian
form  $F$ is \emph{non--degenerate} if and only if $h$ is
non--degenerate. Complete introductions to Hermitian forms over
finite fields may be found in \cite{BC,Se}.

\section{Construction}
\label{s:cons}
 Let  $S$ be a transversal in $\GF(q^2)$ of the
additive subgroup $T_0=\{y\in\GF(q^2): \tr(y)=0\}$, where $\tr: y
\in\GF(q^2) \mapsto y^q+y \in\GF(q)$ is the trace function. Denote by
$\Lambda$ the subset of $GF(q^2)$ satisfying
\begin{equation}\label{prop}
\left(\frac{\alpha-\beta}{\gamma-\beta}\right)^{q-1}\neq 1
\end{equation}
 for any
$\alpha,\beta,\gamma \in\Lambda$.
Choose a basis $B=\{1,\varepsilon\}$ of $GF(q^2)$,
regarded as a $2$--dimensional vector space over $GF(q)$;
hence, it is possible to write each
element $\alpha \in GF(q^2)$ in components
$\alpha_1,\alpha_2\in\GF(q)$  with respect to $B$.
We may thus identify the elements of $GF(q^2)$ with
the points of $AG(2,q)$, by the bijiection
\[ (x,y)\in AG(2,q) \mapsto x+\varepsilon y\in GF(q^2). \]
Condition \eqref{prop} corresponds to require that
$\Lambda$, regarded as point--set in $AG(2,q)$, is an arc. Thus,
setting
 $N=|\Lambda|$, we have
\begin{equation}
\label{eq6}
 N\leq \begin{cases}
q+1&\mbox{ for $q$ odd}\\
   q+2&\mbox{ for $q$ even.\/}
\end{cases}
\end{equation}
%Let $\cF_{\alpha}(x,y)$ be the Hermitian polynomial
Now, consider the  non--degenerate Hermitian forms
$\cF_{\lambda}(X,Y,Z)$ on $GF(q^2)^3$

\[ \cF_{\lambda}(X,Y,Z)= X^{q+1}+Y^{q}Z+YZ^q+\lambda^qX^qZ+\lambda XZ^q, \]
 as $\lambda$ varies in $\Lambda$.
 %order also the elements of $\Lambda$ as $\lambda_1,\ldots,\lambda_N$
%and
%write for any $(x,y)\in\Omega=\GF(q^2)\times S$
%\[ = x^{q+1}+y^{q}+y+\lambda^qx^q+\lambda x \]
%The polynomial $\cF_{\lambda}(x,y)$ is a \emph{Hermitian form} since
%\[\cF_{\lambda}(X,Y)=(X,Y,1) \ H_{\lambda} \begin{pmatrix}
% X^q\\
% Y^q\\
% 1\end{pmatrix}\]
% where  $H_{\lambda}$ is the Hermitian matrix \[H_{\lambda}=\begin{pmatrix}
%    1 &0& \lambda\\
 %   0 & 0& 1\\
 %   \lambda^q &1&0 \\
 % \end{pmatrix}\]
Label the elements of $\Lambda$ as $\lambda_1,\ldots,\lambda_N$
  and let $\Omega=GF(q^2)\times S$.

\begin{theorem}\label{teo}
 The set \[\cC=\{(\cF_{\lambda_1}(x,y,1), \cF_{\lambda_2}(x,y,1),
\ldots,\cF_{\lambda_{N}}(x,y,1))|(x,y)\in    \Omega  \}\]  is a
$q$-ary linear $[N, 3,N-2]$--MDS code.
\end{theorem}
\begin{proof}
%\begin{enumerate}[(a)]
We first  show that $\cC$ consists of $q^3$ tuples from $GF(q)$.
Let $(x_0,y_0), \ (x_1,y_1)\in \Omega $
  and suppose that
  for any $\lambda \in\Lambda$,
  \[ \cF_{\lambda}(x_0,y_0,1)=\cF_{\lambda}(x_1,y_1,1). \]
  Then,
  \begin{equation}\label{eq2}\tr{(\lambda
  (x_1-x_0))}=x_0^{q+1}-x_1^{q+1}+\tr(y_0-y_1).\end{equation}
In particular,
 \begin{equation} \label{eq1}
 \tr{(\lambda (x_1-x_0))}= \tr{(\alpha
 (x_1-x_0))}=\tr{(\gamma
(x_1-x_0))}\end{equation}
 for any $\alpha, \lambda, \gamma \in \Lambda$.

If it were $x_1 \neq x_0$, then \eqref{eq1} would imply
\[\left(\frac{\alpha-\beta}{\gamma-\beta}\right)^{q-1}= 1,\]
contradicting the assumption made on $\Lambda$. Therefore,
$x_1=x_0$ and from \eqref{eq2} we get $\tr(y_0-y_1)=0$. Hence,
$y_0$ and $y_1$ are in the same coset of $T_0$; by definition of
$S$, it follows that $y_0=y_1$, thus $\cC$ has as many tuples as
$|\Omega|$.
%\item

We are now going to show that $\cC$  is a vector subspace of
$\GF(q)^{N}$. Take $(x_0,y_0), (x_1,y_1)\in\Omega$. For any
    $\lambda\in\Lambda$,
    \begin{equation}
      \cF_{\lambda}(x_0,y_0,1)+\cF_{\lambda}(x_1,y_1,1)=
      \cF_{\lambda}(x_2,y_2,1),
    \end{equation}
    where $x_2=x_0+x_1$ and $y_2=y_0+y_1-x_0^qx_1-x_1^qx_0$.
    Likewise, for any $\kappa\in\GF(q)$,
    \begin{equation}
      \kappa\cF_{\lambda}(x_0,y_0,1)=\cF_{\lambda}(x,y,1),
    \end{equation}
    where $x=\kappa x_0$ and $y$ is a
    root of
    \[ y^2+y=(\kappa-\kappa^2)x_0^{q+1}+\kappa(y_0^q+y_0). \]
%\end{enumerate}
Therefore, $\cC$ is a  vector subspace of $\GF(q)^{N}$; as it
consists of $q^3$ tuples,  $\cC$ is  indeed a $3$--dimensional vector
space.

Finally we prove that the minimum distance $d$ of $\cC$ is $N-2$.
Since $\cC$ is a vector subspace of $\GF(q)^{N}$,
 its minimum  distance is  $N-z$, where
  \[ z=\max_{\begin{subarray}{c}
      \mathbf{c}\in\cC\\
      \mathbf{c}\neq\mathbf{0}
      \end{subarray}}|\{i:c_i= 0\}|. \]
    First observe that  $z\geq 2$ because of Singleton bound
    (\ref{Sing1}).
    In order to show that $z=2$ we study the following system
  %   On the other hand,
   % if it were $z\geq 3$, then there uld be
   % at least one
   % $\mathbf{x}\in\Omega$ with $\mathbf{x}\neq (0,0)$
   % such that
    \begin{equation}
      \label{eq:s1}
      \begin{cases}
      \cF_{\alpha}(x,y,1)=0 \\
      \cF_{\beta}(x,y,1)=0 \\
      \cF_{\gamma}(x,y,1)=0 \\
      \end{cases}
    \end{equation}
        for $\alpha,\beta,\gamma$ distinct elements of $\Lambda$.
        Set $U=x^{q+1}+y^q+y$, $V=x^q$ and $W=x$; then,
        \eqref{eq:s1} becomes
        \begin{equation}
          \label{eq:s2}
          \begin{cases}
            U+\alpha^qV+\alpha W=0 \\
            U+\beta^qV+\beta W=0 \\
            U+\gamma^qV+\gamma W =0
          \end{cases}
        \end{equation}
      Since $\left(\frac{\alpha-\beta}{\gamma-\beta}\right)\neq 1$,
      the only solution of \eqref{eq:s2} is $U=V=W=0$, that is
      $x=0$ and $y+y^q=0$. In particular, there is just one solution
      to \eqref{eq:s1} in $\Omega$, that is ${\bf x}=(0,0)$.
      This implies that  a codeword
      which  has at least three zero components is the zero vector, hence
      $z=2$ and thus the minimum distance of $\cC$ is $N-2$.
\end{proof}
{\bf Example.} When $q$ is odd, a transversal $S$ for $T_0$ is
always given by the subfield $\GF(q)$ embedded in $\GF(q^2)$. In
this case it is then extremely simple to construct the code.  For
$q=5$, a computation using GAP \cite{GAP4},  shows that   in order
for $\Lambda$ to satisfy property \eqref{prop}, we  may take
$\Lambda=
\{\varepsilon^3,\varepsilon^4,\varepsilon^8,\varepsilon^{15},\varepsilon^{16},
\varepsilon^{20}\}$, where $\varepsilon$ is a root of the polynomial
$X^2-X+2$, irreducible over $GF(5)$. The corresponding Hermitian
forms are
\[X^{q+1}+Y^{q}Z+yZ^q+\varepsilon^{15}X^qZ+\varepsilon^3 XZ^q\]
\[X^{q+1}+Y^{q}Z+YZ^q+\varepsilon^{20}X^qZ+\varepsilon^4 XZ^q\]
\[X^{q+1}+Y^{q}Z+YZ^q+\varepsilon^{16}X^qZ+\varepsilon^8 XZ^q\]
\[X^{q+1}+Y^{q}Z+YZ^q+\varepsilon^{3}X^qZ+\varepsilon^{15} XZ^q\]
\[X^{q+1}+Y^{q}Z+YZ^q+\varepsilon^{8}X^qZ+\varepsilon^{16} XZ^q\]
\[X^{q+1}+Y^{q}Z+YZ^q+\varepsilon^{4}X^qZ+\varepsilon^{20} XZ^q\]
%\begin{table}
%\[
%\begin{boxedminipage}{12cm}
%\begin{verbatim}
 %q:=5;
% tr2:=function(x)
 %  return x+x^q;
 %end;;
% Fa:=function(a,X)
 %  return X[1]^(q+1)+tr2(X[2])+tr2(X[1]^q*a);
  %end;;
 %L:=Filtered(GF(q)^2,x->x[1]^2+x[2]^2+x[1]*x[2]=Z(q));;
 %La:=Set(L,x->x[1]+Z(q^2)*x[2]);;
 %S:=Elements(GF(q));
 %W:=Cartesian(GF(q^2),S);;
 %GetRow:=function(X);
  %return List(La,a->Fa(a,X));
  %end;
 %CC:=List(W,c->GetRow(c));;
%\end{verbatim}
%\end{boxedminipage}
% \]
%\caption{Construction of [6,3,4]-MDS code} \label{tab:gap}
%\end{table}
A generator matrix for the  $[6,3,4]$--MDS code obtained applying
Theorem \ref{teo} to these Hermitian forms  is,
\bigskip
\[
G=
\begin{pmatrix}
 1& 1& 1& 1& 1& 1 \\
 0& 1& 0& 2& 1& 2 \\
 0& 0& 1& 2& 2& 1
\end{pmatrix}
\]
\medskip
\par
\begin{remark}
In $PG(2,q^2)$, take the line $\ell_{\infty} : Z=0$ as the line at
infinity.
 Then, in the affine space $AG(2,q^2)=PG(2,q^2)\setminus \ell_{\infty}$, any two Hermitian curves
$V(F_{\lambda})$ have $q^2$ affine points in common,  $q$ of which
  in $\Omega \subset AG(2,q^2)$. Likewise, the full intersection
   \[ \bigcap_{\lambda \in \Lambda} V(F_\lambda) \]
   consists of  the $q$ affine points $\{(0,y)|y^q+y=0\}$,
   corresponding
   to just
   a single point in $\Omega$.
\end{remark}

\par
\begin{remark}
Denote by $A_i$ the number of words in $\cC$ of weight $i$.
Since $\cC$ is an MDS code, we have
\[
A_i=\binom{N}{i}(q-1)\sum_{j=0}^{i-N+2}(-1)^j\binom{i-1}{j}q^{i-j-N+2};
 \]
see \cite{Sta}.
Thus,
\[ \begin{array}{lll}
  A_{N-2}&=&\frac{1}{2}(N^2-N)(q-1) \\
  A_{N-1}&=&Nq^2-(N^2-N)q+N^2-2N \\
  A_N    &=&q^3-Nq^2+\frac{1}{2}\left((N^2-N)q-N^2+3N\right).
\end{array}
\]
\end{remark}
\section{Decoding}
\label{decod}
In this section it will be shown how the code $\cC$ we constructed
may be decoded by geometric means.
% WE DON'T USE THIS YET...
%
%In order to simplify our argument, we shall restrict ourselves
%to the case of $q$ odd in which $|\Lambda|=q+1$.

Our approach is based upon two remarks:
\begin{enumerate}[1.]
\item Any received word $\mathbf{r}=(r_1,\ldots,r_N)$ can be
  uniquely
  represented by a set $\widetilde{\mathbf{r}}$ of $N$ points of
  $\PG(3,q)$
  \[
  \widetilde{\mathbf{r}}=\{(\lambda_i^1,\lambda_i^2,r_i,1):
  \lambda=\lambda_i^1+\varepsilon\lambda_i^2\in\Lambda \}.
  \]
  These points all lie
  on the cone $\Psi$ of basis
  \[ \Xi=\{(\lambda_i^1,\lambda_i^2,0,1):
  \lambda=\lambda_i^1+\varepsilon\lambda_i^2\in\Lambda\} \]
  and vertex $Z_{\infty}=(0,0,1,0)$.
\item The function
  \[\phi_{(a,b)}(x,y,z,t)=
  \left(a^{q+1}+\tr(b)\right)t+\tr((x+\varepsilon y)a)
  \]
  is a homogeneous
  linear form defined over $\GF(q)^4$ for any $a,b\in\GF(q^2)$.
\end{enumerate}
Recall that
the codeword $\mathbf{c}$ corresponding to a  given $(a,b)\in\Omega$
is
\[ \mathbf{c}=\left(
\phi_{(a,b)}(\lambda_1^1,\lambda_1^2,0,1),
\phi_{(a,b)}(\lambda_2^1,\lambda_2^2,0,1),
\ldots,
\phi_{(a,b)}(\lambda_N^1,\lambda_N^2,0,1)
\right);
\]
thus,
$\widetilde{\mathbf{c}}$, the set containing
the points $(\lambda_i^1,\lambda_i^2,c_i,1)$, is
the full intersection of the plane
 $\pi_{a,b}: z=\phi_{(a,b)}(x,y,z,t)$ with the cone
$\Psi$.
\par
It is clear that knowledge of the plane $\pi_{(a,b)}$ is enough to
reconstruct the codeword $\mathbf{c}$. In the presence of errors,
we are looking for the nearest codeword $\mathbf{c}$ to a vector
$\mathbf{r}$; this is the same as to determine the plane
$\pi_{(a,b)}$ containing most of the points of
$\widetilde{\mathbf{r}}$. In order to obtain such a plane, we
adopt the following approach. Assume $\ell$ to be a line of the
plane $\pi_{0,0}:z=0$ external to $\Xi$ and denote by
$\pi_{\infty}$ the plane at infinity of equation $t=0$. For any
$P\in\ell$, let $\widetilde{\mathbf{r}}^P$ be the projection from
$P$ of the set $\widetilde{\mathbf{r}}$ on $\pi_{\infty}$. Write
$\cL_{\mathbf{r}}^P$ for a curve of $\pi_{\infty}$ of minimum
degree containing $\widetilde{\mathbf{r}}^P$. Observe that
$\deg\cL_{\mathbf{r}}^P\leq q+1$ and $\deg\cL_{\mathbf{r}}^P=1$
if, and only if, all the points of $\widetilde{\mathbf{r}}$ lie on
a same plane through $P$, that is $\widetilde{\mathbf{r}}$
corresponds to a codeword associated with that plane passing
through $P$.

We now can apply the following algorithm using, for example,
\cite{GAP4}.
\begin{enumerate}[1.]
\item\label{it:1} Take  $P\in\ell$;
\item\label{it:2} Determine the projection $\mathbf{r}^P$ and
  compute the curve $\cL_{\mathbf{r}}^P$;
\item Factor $\cL_{\mathbf{r}}^P$ into irreducible factors,
  say $\cL_1,\cL_2,\ldots,\cL_v$;
\item Count the number of points in $\widetilde{\mathbf{r}}^P\cap
  V(\cL_i)$ for any factor $\cL_i$ of  $\cL_{\mathbf{r}}^P$ with
  $\deg\cL_i=1$.
\item\label{it:5}
  If for some $i$ we have  $n_i>\frac{N+1}{2}$, then return the plane
  spawned by $P$ and two points of $L_i$;
  else, as long as not all the points of $\ell$ have been
  considered, return to point \ref{it:1}.
\item If no curve with the required property has
  been found, return failure.
\end{enumerate}
\begin{remark}
  The condition on $n_i$ in point \ref{it:5}
  checks if the plane contains more than half of the
  points corresponding to the received word $\mathbf{r}$;
  when this is  the case, a putative codeword $\mathbf{c}$
  is constructed, with
  $d(\mathbf{c},\mathbf{r})\leq\frac{N-3}{2}$;
  thus, when $\mathbf{c}\in\cC$, then it is indeed the
  unique word of $\cC$ at minimum distance from
  $\mathbf{r}$.
  However, the aforementioned algorithm may be altered in
  several ways, in order to be able to try to correct
  errors beyond the bound; possible approaches are
  \begin{enumerate}
  \item iterate the procedure for all the points on $\ell$
    and return the planes containing most of the points
    corresponding to the received vector;
  \item use some further properties of the cone $\Psi$;
    in particular, when $\Xi$ is a conic
    it seems possible to improve the decoding by considering
    also the quadratic components of the curve $\cL_{\mathbf{r}}^P$.
  \end{enumerate}
\end{remark}
\begin{remark}
  The choice of $P$ on a line $\ell$ is due to the fact that
  any line of $\pi_{0,0}$ meets all the planes of $\PG(3,q)$.
  In general, we might have chosen $\ell$ to be just a
  blocking set disjoint from $\Xi$.
  Observe that when $q$ is odd and
  $|\Lambda|=q+1$, then the line $\ell$ is just an external line to
  a conic of $\pi_{0,0}$.
\end{remark}

\bigskip
\begin{minipage}{6.8cm}
\begin{obeylines}
{\sc Angela Aguglia}
 Dipartimento di Matematica
 Politecnico di Bari
 Via G. Amendola 126/B
 70126 Bari
 Italy
 {\tt
a.aguglia@poliba.it}
\end{obeylines}
\end{minipage}
\qquad
\begin{minipage}{6.8cm}
\begin{obeylines}
{\sc Luca Giuzzi}
 Dipartimento di Matematica
 Politecnico di Bari
Via G. Amendola 126/B
70126 Bari
Italy
{\tt l.giuzzi@poliba.it}
\end{obeylines}
\end{minipage}

\end{document}